# Enterprise Content Management: Theory and Engineering for Entire Lifecycle Support

Dr. Sergey V. Zykov, Ph.D.
ITERA Oil and Gas Company L.L.C.
Moscow, Russia
e-mail: szykov@itera.ru

**Abstract**[1]

The paper considers enterprise content management (ECM) issues in global heterogeneous distributed computational environment. Present-day enterprises have accumulated a huge data burden. Manipulating with such a bulk becomes an essential problem, particularly due to its global distribution, heterogeneous and weak-structured character. The conceptual approach to integrated ECM lifecycle support is presented, including overview of formal models, software development methodology and innovative software development tools. Implementation results proved shortening terms and reducing costs of implementation compared to commercial software available.

## 1. Introduction

Due to the tendency of almost annual doubling of enterprise information size, state-of-the-art information systems (IS) and DBMS are currently operating terabyte information arrays, which are going to change for petabyte size shortly.

Tremendous heterogeneous data volumes demand new software development tools that provide continuous iterative design, implementation and support for IS integrating both DB and meta-DB (MDB) containing specific information for DB handling. Since contemporary scholars use MDB for knowledge acquisition and processing, solving problem of integrated DB and MDB management by means of ECM software provides potentially new level of information handling, thanks to data and knowledge cooperation.

Under rapid IT penetration into each and every activity sphere of the modern society, software integration through ECM becomes a critical issue, particularly considering DB and MDB sizes and co-existing heterogeneous (and often contradicting) concepts, methodologies, models and approaches to handling them. The fact that major software development companies (such as Microsoft, IBM, Oracle, SAP, BEA etc.) have not produced a uniform approach to complex IS construction yet (including Internet-based ones) speaks for itself; universal terminology has not been worked out either. Quite a number of research groups focus on a uniform theoretical language and SDK solution for enterprise portal design, implementation and integration. However, the problem is still far from its adequate solution.

The above considerations require a conceptually new approach to portal-integrated IS design, implementation and support, which provides dynamic tracking (meta)data integrity and high IS fault tolerance in heterogeneous computational environment.

## 2. The Methodology

### 2.1. Overview

The suggested conceptual approach to ECM-based data lifecycle support is based on (meta)data integration and aimed primarily to conquer the obstacles mentioned.

One of its essential components is conceptual software design, which provides a unified (meta)data computational model in the form of (meta)data objects, language and software development tools for manipulating them.

Another essential ingredient of the approach is the methodology that supports continuous ECM iterative software design and implementation (including reengineering) from problem domain entities to target enterprise IS schemes. Therewith, (meta)data actuality, completeness, consistency and integrity is controlled throughout the entire enterprise content lifecycle.

One more innovative aspect of the research deals with IS and (M)DB CASE-, RAD-, and integration tools. Application development requires conceptual and methodological generalization of (M)DO management processes to provide integrated DO and MDO management on the basis of uniform, open and extendable interface, language and software tools.

The research methods are based on a creative synthesis of fundamental statements of finite sequence theory [1], category theory [2,3], computations theory [5] and semantic networks theory [4].

Below, an ECM design methodology is presented, which supports the entire lifecycle. The approaches known as yet either have methodological "gaps" or do not result in enterprise-level solutions with practically applicable implementation features (scalability, expandability, availability, fault tolerance etc.).

The developed data models (DM) for problem domains (variable domain-based) and software development tools (abstract machine-based) manipulate dynamic and static features of heterogeneous weak-structured environments in a more detailed way.

The variable domain-based DM with states features event-driven DO and MDO control of heterogeneous problem domains. Therewith, the range of possible (meta)data sources is extended up to virtually arbitrary data warehouses (including (M)DB), that support state-of-the-art front-end architectures of globally distributed IS, intermediate and legacy systems. The ECM lifecycle solution features content-oriented (M)DM, which is modeled by an abstract machine. As far as implementation part is concerned, traditional technologies and software development tools are extended by conceptual integrated ECM design based on a creative combination of object models, UML language and BPR technology applied to a decentralized, personalized, globally distributed component environment.

Research attention is focused on object-oriented portal design and implementation that provides front-end interfaces with heterogeneous enterprise data warehouses by built-in procedures and embedded dynamic modules.

## 2.2. Enterprise Content Definition Model

Additionally, if you have any suggestions as to how the instructions could be made clearer or easier to use, please let us know.

The computational data model (CDM) is suggested in a form of object calculus.

DO of the CDM can be represented as follows:

*DO* = <*class, object, value*>,

where under a *class* a collection of problem domain objects is implied. An *object* a class instantiation by the enterprise CMS, which partially evaluates the class metadata. A *value* means an HTML web-page entity, which is generated by the enterprise CMS so that the metadata of the template is fully evaluated. *State* changes model problem domain individual dynamics.

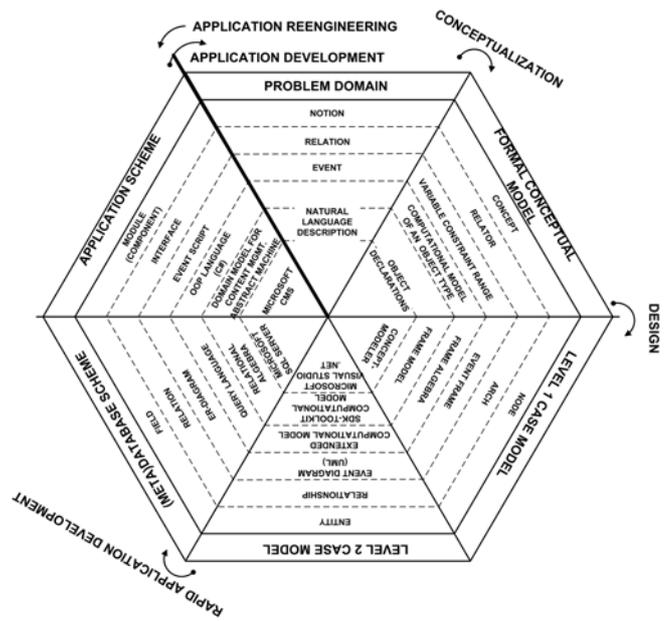

**Figure 1. General ECM scheme**

Compared to results obtained as yet, principal benefits of the CDM suggested are more adequate mapping of heterogeneous weak-structured problem domain dynamics and statics as well as event-driven (meta)data control in a global computational environment. As to architecture and interface, the CDM provides "penetrating" iterative design of open, distributed, interoperable portal based on UML and BPR methodologies and .NET web services. The implementation supports an integrated front-end information management for various data warehouse types of heterogeneous enterprise IS on the technological basis of event-driven procedures and dynamic SQL-based component modules.

The globally distributed information management problem domain requires support for multi-level profile-based access to heterogeneous distributed databases and data warehouses based on unified Internet information interchange protocols and web services. To meet the ECM interface requirements, dynamically variable (meta)data representation, flexible personalized access granting and consistent support of data integrity should be provided. The ECM system architecture should be open, extendable, flexibly adaptable to problem domain state; it should also provide online data and metadata personalized correction (i.e. depend on the ECM environment state).

The globally distributed information management problem domain requires support for multi-level profile-based access to heterogeneous distributed databases and data warehouses based on unified Internet information interchange protocols and web services. To meet the ECM interface requirements, dynamically variable (meta)data representation, flexible personalized access granting and consistent support of data integrity should be provided. ECM system architecture should be open, extendable, flexibly adaptable to problem domain state; it should also provide online data and metadata personalized correction (i.e. depend on the ECM environment state).

A multi-level ECM system design and implementation scheme is suggested (fig. 1) that provides fast component-based development of integrated, open, extendable IS for global networks with continuous (meta)data adequacy and integrity control. During the development procedure IS specification is transformed from problem domain notations to formal CDM entities, further, by CASE-toolkit to OR(M)DB scheme with an ECM system (e.g. Oracle Portal) as (M)DO manipulation tool, and, finally, to target IS formal architecture and interface component description.

The developed computational model is based on two-level *conceptualization* [6], i.e., the process of establishing relations between problem domain concepts.

Objects $o$, according to assigned types $T$, are assembled into assignment-dependent collections, thus forming variable domains

$O_T(A) = \{o | o : A \to T\}$,

which model problem domain dynamics and statics.

When fixing CDM individuals, uniqueness of DO $d$ individualization within problem domain $D$ by formula $\Delta$ should be maintained:

$||Iv\ \Delta(v)||i = d \Leftrightarrow \{d\} = \{d \in D\ |\ ||\Delta(d)||i=1\}$.

Computational model semantics for (meta)data and states are adequately and uniformly formalized by multi-sort typed $\lambda$-calculus [1] and combinatory logic [3], semantic network-based situational description [4] and state-based abstract (virtual) categorical machine [2].

The DO model compression principle

$C = Iw : [D]\ \forall v : D(w(v) \spadesuit \Delta) = \{v: D\ |\ \Delta\}$

allows model application to classets, objects and values separately and to the DO on the whole.

Metadata computational model extends Codd's relational model by the compression principle:

$v^{j+1} \equiv I\ w^{j+1}:[...[D]...]\ \forall v^j:[...[D]...](w^{j+1}(v^j) \leftrightarrow \Delta)$

where

$w^{j+1}, z^{j+1}$ are metalevel predicate symbols in relation to level $j$;

$v^j$ is an individual of level $j$;

$\Delta^j$ is a DO definition language construct of level $j$.

The above integrated object model for data, metadata and states is characterized by structural hierarchical organization, scalability, metadata encapsulation and readability. Extendibility, adequacy, neutrality and semantic soundness of the formalization provide problem-oriented IS development with (M)DO adequacy maintenance throughout the entire lifecycle.

During the research, automated translation procedure is developed that transforms (M)DO of the above CDM into target (M)DB schemes and (meta)data management abstract machine codes; the machine models ECM.

On the basis of multi-parameter functional

$F = F\ ((v), (e), ...) (s) (p)$,

where assignment values represent:

$s$ – IS user personal preferences;

$p$ – IS user registration status;

$v$ – IS client interface parameters;

$e$ – IS data access device parameters,

semantics object model and formal generalized procedure of (meta)data instantiation have been built, depending on the above assignments and based on functional $F$ evaluation function $||\circ||$ [8].

## 2.3 Content Management Model

Abstract machine for content management (AMCM) [9,10] is suggested as a formal model of portal-based content management, which is an improved version of categorical abstract machine (CAM) [2]. At any given moment AMCM is determined by its *state*. AMCM work *cycle* can be formalized by explicit enumeration of possible state changes, which define the procedure of AMCM state *dynamics* modeling.

From the formal model viewpoint, when portal page templates are mapped into the pages, variable *binding* actually occurs, which is evaluation of variables that characterize template elements with their values, or portal page elements.

AMCM semantics can be described on the basis of D.Scott semantic domain theory [5]. Therewith, atomic template types are chosen out of standard domains, while more complex template types are built using domain constructors.

AMCM formal semantics is built in the following order:

1. *Standard* (most commonly used within the model framework) domain enumeration;

2. *Finite* (containing explicitly enumerable elements) domain definition;

3. Domain *constructor* (operations of building new domains out of the existing ones) definition, which are domain combination operators;

4. Aggregate domain formalization using standard domains and domain constructors.

Domain constructors include functional space $[D_1 \rightarrow D_2]$, Cartesian product $[D_1 \times D_2 \times ... \times D_n]$, sequence $D^*$ and disjunctive sum $[D_1 + D_2 + ... + D_n]$.

Let the AMCM language contain expression set $E$ (including constant set, identifier set $I$, assignment operation (content "write operation" to template "slot") etc.), and command set $C$ (comparison, command sequence etc.).

AMCM syntax is completely defined by the following syntax domain description:

*Ide* = {*I* | *I* – *identifier*};

*Com* = {*C* | *C* – *command*};

*Exp* = {*E* | *E* – *expression*}.

Let us collect all possible language identifiers into *Ide* domain, commands – into *Com* domain, and expressions – into *Exp* domain.

State-based computational model of AMCM environment is as follows:

*St* = *Mem* × *In* × *Out*;

*Mem* = *Id* → [*Val* + {*unbound*}];

*In* = *Val**;

*Out* = *Val**;

*Val* = *Type1* + *Type2* + ...

AMCM state is defined by "memory" state considering input values (i.e. content) and output values (i.e. web-pages) of the abstract machine. Therewith, under *memory* a mapping from identifier domain into value domain is implied, which is similar to lambda-calculus variable binding. For correct exception handling, *unbound* element should be added to the domains. *Value* domain is formed by disjunctive sum of domains, which contain content types of AMCM language.

Semantic statements describe *denotates* (i.e. correct construct values) of AMCM (M)DO manipulation language.

Semantic statement for an expression reads:

***E***: *Exp* → [ *State* → [[*Value* × *State*] + {*error*}]];

Expression evaluation in AMCM environment results in such a state change that the variable is bound to its value, or (in case the binding is impossible due to variable and value type incompatibility) an error is generated.

Semantic statement for a command reads:

***C*** : *Com*→[*State*→[*State*+{*error*}]].

Generally speaking, expression evaluation in AMCM environment results in AMCM state change and a situation is possible (e.g., assignment type incompatibility), when an error is generated.

Constant denotates are their respective values (in a form of ordered pair of <*variable, value*>), while program state remains unchanged.

Identifier denotates are identifiers bound with their values in case binding is possible (in a form of ordered tuples of <*variable_in_memory, identifier, state*>), while the state remains unchanged, and an error message is generated in case the binding is impossible:

***E*** [*I*] *s* = (*m*, *I* = *unbound*) *error*, →(*m*, *I*, *s*).

Semantic function for assignment command of AMCM language has the following type:

***C***: *Com* → *State* → [*State* + {*error*}].

Thus, content management IS template binding with the content may result in AMCM state change and in a number of limited, predefined cases (particularly, under template and content type incompatibility) – in error generation.

Semantic statement for an AMCM command, which assigns content to template element, results in state change with substitution of content value *v* by identifier *I* in the memory:

***C*** [*I=E*] = ***E*** [*E*] * λ*v* (*m* , *i*, *o*) . (*m* [*v*/*I*], *i*, *o*).

## 3. From Model to Implementation

Portal development concepts and methodology have been instantiated for enterprise resource management, including information and interface (i.e. data and metadata) of enterprise information resources (Internet and Intranet sites). Detailed design scheme consists of the following stages (Zykov, 2004b):

1. Management board outlines objectives and measures to control information resources that map into formal business rules of portal computational model;

2. Information resource managers build detailed structure and functional enterprise conceptual business model as a DO map;

3. System analysts consider OLAP-instantiations of the enterprise business model versions for various development scenarios;

4. Portal and DB developers formalize architecture and interface logics with OO scripting languages (in terms of state-based abstract machine), which is later translated into UML-based DM (abstract machine code) by synthetic CASE methodology;

5. DB, LAN, Internet, Intranet and security administrators, designers and content managers implement and support target configurations for the portal, DB and information network resources.

According to detailed development sequence, a generalized heterogeneous data warehouse processing procedure is suggested that allows users to interact with distributed (M)DB in a certain state, depending upon dynamically script-activated assignments. Therewith, scripts are initiated (in the form of (M)DB connection profiles and state-based abstract machine OO script procedures) depending on user-triggered events, which provide transparent intellectual client-server front-end interaction. Dynamically variable profiles for information resource and interface (M)DB access provide strict and flexible personalization, high fault tolerance and data security for ordinary and privileged portal users.

## 4. Implementation Features

The suggested conceptual approach to portal design and implementation has been practically approved by development of Internet and Intranet portals in ITERA International Group of Companies.

As to information management software for global computational environment, its *Menu* component is aimed at portal navigation data storage and processing. *Pages* subsystem is related to *Menu* and tracks events of assignment, transfer, and deletion of portal pages (meta)data and navigation menu items. *Images* module provides storage, retrieval and portal web publication of digitized photos and graphics. *News Columns* supports periodical portal publications (press releases, media news etc.), including data from related third-party modules. *Special sections* organizes visual management of portal content (i.e. data) and design (i.e. metadata) by given criteria set. *Administration* implements profiling, data access policies and (M)DB synchronization.

In terms of system architecture, the portal provides assignments (depending on front-end position in data access hierarchy) with a certain level of (meta)data entry, modification, analysis and report generation (from administrator down to reader). Problem-oriented form designer, report writer, online documentation and administration tools make an interactive interface toolkit. (M)DB supports integrated storage of data (for online access) and metadata (DO dimensions, integrity constraints, data representation formats and other information resource parameters).

During the resource management IS design, problem domain DM specifications (in terms of semantic networks) are transformed by *ConceptModeller* to UML diagrams, then by Oracle Developer/2000 integrated CASE tool – to ER diagrams (or by AMCM and Oracle Portal toolkit – into AM code) and, finally, into target IS and (M)DB schemes.

Using the suggested information model, the architectural and interface solution has been customized for resource management IS with (M)DB interaction assignments for various classes of users and administrators.

Portal implementation process included fast prototyping (with SQL and Perl stored OO scripts) and full-scale integrated Oracle-based implementation.

The fast portal prototype has been designed to prove adequacy of the formal (meta)data models, methods and algorithms, it uses a generalized architecture scheme and linking interfaces. As a result of comparative analysis the environment has been chosen that included Sybase S-Designor and PowerBuilder CASE-and-RAD toolkit, Perl language and mySQL DBMS.

Upon prototype testing, a full-scale enterprise OO toolkit has been implemented. Web pages automatically generated by information resources (meta)data management system are published at ITERA Group Intranet portal and official Internet site (www.itera.ru).

To provide the required industrial scalability and fault tolerance level, the integrated Oracle design and implementation toolkit (Portal, Developer/2000) has been chosen to support UML and BPR methodologies.

All of the components are designed, implemented and customized according to technical specifications outlined by the author and tested for several years in a heterogeneous enterprise environment. The resulting implementation terms and costs have been reduced about 40% (on the average) compared to commercial software available, while functional characteristics have been essentially improved. Advanced personalization and access level differentiation allows to substantially reduce risks of (meta)data damage or loss (Zykov, 2003).

## 5. Conclusion

A comprehensive conceptual approach to portal-based IS design, implementation and maintenance has been outlined. Based on the approach, a new methodology has been developed for information resource management, which provides adequate, consistent and integrate (meta)data manipulation during the entire lifecycle (Zykov, 2002).

Upon customizing theoretical methods of finite sequences, categories, semantic networks, computations and abstract machines, a set of models have been constructed including problem domain conceptual model for (M)DO dynamics and statics as well as a model for development tools and computational environment in terms of state-based abstract machines, which provide integrated (M)DO manipulation in (weak-structured) heterogeneous problem domains. For the model collection, a generalized development toolkit choice criteria set has been suggested for IS prototyping, design and implementation.

A SDK has been implemented including *ConceptModeller* visual problem oriented CASE-tool and the application for content (i.e., (M)DO) management.

According to the conceptual approach, a generalized interface solution has been designed for Internet-portal, which is based on content-oriented architecture with explicit division into front-end and back-end sides.

To solve multi-aspect applied task of information resource management, a fast event-driven prototype has been developed using *ConceptModeller* toolkit and PowerScript and Perl programming languages. After the prototype testing, a full-scale enterprise object-oriented IS has been designed and implemented as an information resource management application.

The full-scale enterprise portal has been customized for information resource management and implemented in a corporation employing nearly 10,000 people. The obtained results proved shortening terms and reducing costs of implementation compared to commercial software available, and revealed high mobility, expandability, scalability and ergonomics. Portal design scheme is based on a formal model synthesizing OO methods of management of DO (i.e. data) and MDO (i.e. knowledge).

Practical value of the results obtained is defined by the benefits of large-scale portal development by the methods suggested. Particularly, formal DM is used, which synthesizes methods of finite sequences, category theory, computation theory and semantic networks and that provides joint DO and MDO management in heterogeneous interoperable globally distributed environment. Due to formalized testing and verification procedures, costs of (meta)data maintenance, fault tolerance and integrity support have been essentially reduced, while portal modernization, customization and performance optimization procedures have been simplified.

The results obtained have been used for development of a number of portal in ITERA Group:

- Content management system for network information resources (2001-2002);
- Official Internet site, www.iteragroup.com (2003-2004);
- Enterprise Intranet portal (2004).

Models, methods and tools developed by the author served as a basis for portal-based distributed information resource management in ITERA International Group of Companies. According to ITERA experts, the portal implementation has resulted in annual cost reduction of hundreds of thousands of US dollars, while data management efficiency has increased substantially.

The author is going to continue his studies of the formal models and related SDK that support enterprise portals.